# The Dividend Discount Model with Multiple Growth Rates of Any Order for Stock Evaluation

Abdulnasser Hatemi-J and Youssef El-Khatib

UAE University, Al-Ain, P.O.Box 15551, United Arab Emirates.

E-mails: AHatemi@uaeu.ac.ae and Youssef_Elkhatib@uaeu.ac.ae

Abstract

In this paper we provide a general solution for the dividend discount model in order to compute the intrinsic value of a common stock that allows for multiple stage growth rates of any predetermined number of periods. A mathematical proof is provided for the suggested general solution. A numerical application is also presented. The solution introduced in this paper is expected to improve on the precision of stock valuation, which might be of fundamental importance for investors as well as financial institutions.

## 1. Introduction

Determining a measure that can represent the intrinsic value of a stock is an important issue for investors and financial institutions seeking profitable investment prospects. This issue might be also of interest for policy makers in order to design appropriate financial guidance. The dividend discount model (DDM), which has originally been developed by Gordon and Shapiro (1956) and Gordon (1959, 1962), can be used for this purpose.[1] Currently, there are extensions of the model in the literature that allow for the valuation of a common stock with two different growth rates across time at the maximum, to our best knowledge.[2] In this paper we suggest a general solution for the valuation of common stocks for the DDM that allows for multiple growth rates of any predetermined number. The suggested solution is proved mathematically and an application is provided.

The remaining part of this paper is organized as follows. The next section presents the model and provides a general solution along with a mathematical proof. Section 3 applies the suggested solution to valuate a common stock with three different growth rates. The last section offers the concluding remarks.

## 2. The Model with Multiple Growth Rates for Stock Evaluation

We consider $N$ different growth annuity periods $T_1$ with growth $g_1$, $T_2$ with growth $g_2$, $\cdots$, and $T_N$ with growth $g_N$, followed by a perpetuity with growth rate $g_{N+1}$. For $1 \leq l \leq N$, we denote by $ST_l := \sum_{i=0}^{i=l} T_i$ the end of period $T_l$. The time is to be going from $t = T_0 := 0$ to $ST_N := \sum_{i=0}^{i=N} T_i$ with $N$ annuities and from $ST_N + 1$ to $\infty$ with a perpetuity as explained in the following:

During $T_0 \to T_1 = ST_1$    the growth rate is $g_1$,

The dividend at year $i$ before $ST_1$ is      $D_i = D_0(1+g_1)^i$

---

[1] See also Gordon and Gould (1978) as well as Fuller and Hsia (1984).
[2] It should be mentioned that there are also alternative models for stock valuation such as the capital asset pricing model developed by Sharpe (1964) and the arbitrage pricing theory suggested by Ross (1976).

The dividend at the end of the period is $D_{ST_1} = D_0(1+g_1)^{T_1}$

During $T_1 + 1 = ST_1 + 1 \to T_1 + T_2 = ST_2$ the growth rate is $g_2$,

The dividend at year $i$ between $ST_1$ and $ST_2$ is $D_i = D_{ST_1}(1+g_2)^i$

The dividend at the end of the period is $D_{ST_2} = D_{ST_1}(1+g_2)^{T_2}$

Continuing recusively until the following:

During $ST_{k-1} + 1 \to ST_k$ the growth rate is $g_k$,

The dividend at year $i$ between $ST_{k-1}$ and $ST_k$ is $D_i = D_{ST_{k-1}}(1+g_k)^i$

The dividend at the end of the period is $D_{ST_k} = D_{ST_{k-1}}(1+g_k)^{T_k}$

And similarly by continuing we have

During $ST_{N-1} + 1 \to ST_N$ the growth rate is $g_N$,

The dividend at year $i$ between $ST_{N-1}$ and $ST_N$ is $D_i = D_{ST_{N-1}}(1+g_N)^i$

The dividend at the end of the period is $D_{ST_N} = D_{ST_{N-1}}(1+g_N)^{T_N}$.

Moreover, for any $1 \leq k \leq N$ and $0 \leq l \leq T_{k+1}$ we can expresse the following:

$$D_{ST_k+l} = D_{ST_k}(1+g_{k+1})^l = \left[D_0 \prod_{i=0}^{k}(1+g_i)^{T_i}\right](1+g_{k+1})^l. \tag{1}$$

The dividend after the $ST_N + m$ years is $D_{ST_N+m} = D_{ST_N}(1+g_{N+1})^m$.

Now we can present our main result via the following proposition.

**Proposition 1** The value of a share, $P_0$, is given by

$$P_0 = \sum_{k=1}^{N+1} \frac{D_{ST_{k-1}+1}}{(1+r)^{ST_{k-1}}(r-g_k)} A_k, \tag{2}$$

where

$$A_k := \left[1 - \left(\frac{1+g_k}{1+r}\right)^{T_k}\right], \quad \text{for any } 1 \leq k \leq N, \tag{3}$$

$$A_{N+1} := 1. \tag{4}$$

*Proof.* First, notice that the value of a share is given -in general terms- by the series

$$P_0 = \sum_{i=1}^{\infty} \frac{D_i}{(1+r)^i}, \tag{5}$$

where $D_i$ denotes the dividend at year $i$. In the case of $N$ annuities periods $T_1, \ldots, T_N$ with respective growth rate $g_1, \ldots, g_N$ followed by a perpetuity with growth rate $g_{N+1}$ the value of a share can be written as

$$P_0 = \sum_{i_1=1}^{T_1} \frac{D_0(1+g_1)^{i_1}}{(1+r)^{i_1}} + \sum_{i_2=T_1+1}^{T_1+T_2} \frac{D_0(1+g_1)^{T_1}(1+g_2)^{i_2-T_1}}{(1+r)^{i_2}} + \cdots$$

$$+ \sum_{i_k=ST_{k-1}+1}^{ST_k} \frac{D_{ST_{k-1}}(1+g_k)^{i_k-ST_{k-1}}}{(1+r)^{i_k}} + \cdots + \sum_{i_N=ST_{N-1}+1}^{ST_N} \frac{D_{ST_{N-1}}(1+g_N)^{i_N-ST_{N-1}}}{(1+r)^{i_N}}$$

$$+ \sum_{i_{N+1}=ST_N+1}^{\infty} \frac{D_{ST_N}(1+g_{N+1})^{i_{N+1}-ST_N}}{(1+r)^{i_{N+1}}}.$$

Then,

$$P_0 = \sum_{k=1}^{N} \sum_{i_k=ST_{k-1}+1}^{ST_k} \frac{D_{ST_{k-1}}(1+g_k)^{i_k-ST_{k-1}}}{(1+r)^{i_k}} + \sum_{i_{N+1}=ST_N+1}^{\infty} \frac{D_{ST_N}(1+g_{N+1})^{i_{N+1}-ST_N}}{(1+r)^{i_{N+1}}}$$

$$= \sum_{k=1}^{N} D_{ST_{k-1}} \sum_{l=1}^{ST_k-ST_{k-1}} \frac{(1+g_k)^l}{(1+r)^{l+ST_{k-1}}} + D_{ST_N} \sum_{j=1}^{\infty} \frac{(1+g_{N+1})^j}{(1+r)^{j+ST_N}}$$

$$= \sum_{k=1}^{N} \frac{D_{ST_{k-1}}}{(1+r)^{ST_{k-1}}} \sum_{l=1}^{T_k} \left[\frac{1+g_k}{1+r}\right]^l + \frac{D_{ST_N}}{(1+r)^{ST_N}} \sum_{j=1}^{\infty} \left[\frac{1+g_{N+1}}{1+r}\right]^j. \tag{6}$$

The last line contains two sums of geometric series that can be calculated as follows

$$\sum_{l=1}^{T_k}\left[\frac{1+g_k}{1+r}\right]^l = \sum_{l=0}^{T_k}\left[\frac{1+g_k}{1+r}\right]^l - 1$$

$$= \frac{1-\left(\frac{1+g_k}{1+r}\right)^{T_k+1}}{1-\frac{1+g_k}{1+r}} - 1 = \frac{1+g_k}{1+r}\left[\frac{1-\left(\frac{1+g_k}{1+r}\right)^{T_k}}{1-\frac{1+g_k}{1+r}}\right] = \frac{1+g_k}{r-g_k}\left[1-\left(\frac{1+g_k}{1+r}\right)^{T_k}\right], \quad (7)$$

and

$$\sum_{j=1}^{\infty}\left[\frac{1+g_{N+1}}{1+r}\right]^j = \sum_{j=0}^{\infty}\left[\frac{1+g_{N+1}}{1+r}\right]^j - 1 = \frac{1}{1-\frac{1+g_{N+1}}{1+r}} - 1 = \frac{1+g_{N+1}}{r-g_{N+1}}. \quad (8)$$

Substituting equations (7) and (8) in (6) we obtain

$$P_0 = \sum_{k=1}^{N}\frac{D_{ST_{k-1}}}{(1+r)^{ST_{k-1}}}\frac{1+g_k}{r-g_k}\left[1-\left(\frac{1+g_k}{1+r}\right)^{T_k}\right] + \frac{D_{ST_N}}{(1+r)^{ST_N}}\frac{1+g_{N+1}}{r-g_{N+1}},$$

$$P_0 = \sum_{k=1}^{N+1}\frac{D_{ST_{k-1}+1}}{(1+r)^{ST_{k-1}}(r-g_k)} A_k,$$

where $A_k$, for $1 \leq k \leq N+1$, are given by equations (3) and (4). This ends the underlyoing proof.

In the previous proposition, we provide the value of a share as a function of different growth rates $g_i$ and dividends $D_{ST_i}$ at end of period $ST_i$, for the interest rate $r$. We can derive a more explicit formula as a function of $D_0$, with different growth rates $g_i$ and the interest rate $r$ as is shown in the following corollary.

**Corollary 1** *The value of a share $P_0$ is given by*

$$P_0 = D_0 \sum_{k=1}^{N+1}\frac{\prod_{i=0}^{k-1}(1+g_i)^{T_i}(1+g_k)}{(1+r)^{ST_{k-1}}(r-g_k)} A_k, \quad (9)$$

where $A_k$, for $1 \leq k \leq N+1$, are given by equations (3) and (4).

*Proof.* The proof is straightforward by embedding the value of $D_{ST_{k-1}+1}$ from (1) into the formula (2).

**Example 1** In this example, we consider the case of $N = 1$, i.e. when there is one growth rate $g_1$ at time $T$ and after that a perpetuity with growth rate $g_2$. The time is then going from $T_0 = 0$ to $T_1 = T$ with a growth rate $g_1$ and after $T$ with growth rate $g_2$. The formula (9) for this particular case can be written as

$$P_0 = \frac{D_0(1+g_1)}{(r-g_1)} A_1 + \frac{D_0(1+g_1)^{T_1}(1+g_2)}{(1+r)^{ST_1}(r-g_2)} A_2 = \frac{D_1}{(r-g_1)}\left[1-\left(\frac{1+g_1}{1+r}\right)^T\right] + \frac{D_{T+1}}{(1+r)^T(r-g_2)}.$$

**Example 2** In this example, we consider the case of $N = 2$, i.e. we have annuities with growth rates $g_1$ for time $T_1$, $g_2$ for time $T_2$ and after that a perpetuity with growth rate $g_3$. The time is then going from $T_0 = 0$ to $T_1$ with a growth rate $g_1$, from $T_1+1$ to $T_1+T_2$ with a growth rate $g_2$ and after $T_1+T_2$ with the growth rate $g_3$. Using formula (2) and equations (3-4) we have

$$P_0 = \sum_{k=1}^{3} \frac{D_{ST_{k-1}+1}}{(1+r)^{ST_{k-1}}(r-g_k)} A_k$$

$$= \frac{D_{ST_0+1}}{(1+r)^{ST_0}(r-g_1)} A_1 + \frac{D_{ST_1+1}}{(1+r)^{ST_1}(r-g_2)} A_2 + \frac{D_{ST_2+1}}{(1+r)^{ST_2}(r-g_3)} A_3,$$

where

$$ST_0 = 0, \ ST_1 = T_1, \ \text{and} \ ST_2 = T_1 + T_2,$$

and

$$A_1 = \left[1-\left(\frac{1+g_1}{1+r}\right)^{T_1}\right], \ A_2 = \left[1-\left(\frac{1+g_2}{1+r}\right)^{T_2}\right], \ A_3 = 1.$$

Thus, the following can be expressed:

$$P_0 = \frac{D_1}{(r-g_1)}\left[1-\left(\frac{1+g_1}{1+r}\right)^{T_1}\right] + \frac{\frac{D_{T_1+1}}{(r-g_2)}}{(1+r)^{T_1}}\left[1-\left(\frac{1+g_2}{1+r}\right)^{T_2}\right] + \frac{\frac{D_{T_1+T_2+1}}{(r-g_3)}}{(1+r)^{T_1+T_2}}.$$

**Example 3** In this example, we consider the case of $N = 3$ i.e. we have annuities with growth rates $g_1$ for time $T_1$, $g_2$ for time $T_2$, $g_3$ for time $T_3$ and after that a perpetuity with the growth rate $g_4$. Again, using formula (2) and equations (3-4) we have

$$P_0 = \sum_{k=1}^{4} \frac{D_{ST_{k-1}+1}}{(1+r)^{ST_{k-1}}(r-g_k)} A_k$$

$$= \frac{D_{ST_0+1}}{(1+r)^{ST_0}(r-g_1)} A_1 + \frac{D_{ST_1+1}}{(1+r)^{ST_1}(r-g_2)} A_2$$

$$+ \frac{D_{ST_2+1}}{(1+r)^{ST_2}(r-g_3)} A_3 + \frac{D_{ST_3+1}}{(1+r)^{ST_3}(r-g_4)} A_4,$$

where

$$ST_0 = 0, \ ST_1 = T_1, \ ST_2 = T_1 + T_2, \ ST_3 = T_1 + T_2 + T_3$$

and

$$A_1 = \left[1 - \left(\frac{1+g_1}{1+r}\right)^{T_1}\right], \ A_2 = \left[1 - \left(\frac{1+g_2}{1+r}\right)^{T_2}\right], \ A_3 = \left[1 - \left(\frac{1+g_3}{1+r}\right)^{T_3}\right], \ A_4 = 1.$$

Thus, we have

$$P_0 = \frac{D_1}{(r-g_1)}\left[1 - \left(\frac{1+g_1}{1+r}\right)^{T_1}\right] + \frac{\frac{D_{T_1+1}}{(r-g_2)}}{(1+r)^{T_1}}\left[1 - \left(\frac{1+g_2}{1+r}\right)^{T_2}\right]$$

$$+ \frac{\frac{D_{T_1+T_2+1}}{(r-g_3)}}{(1+r)^{T_1+T_2}}\left[1 - \left(\frac{1+g_3}{1+r}\right)^{T_3}\right] + \frac{\frac{D_{T_1+T_2+T_3+1}}{(r-g_4)}}{(1+r)^{T_1+T_2+T_3}}.$$

## 3. An Application

Suppose a bank has just paid a dividend per share of $2. Its dividend is expected to grow at 5% during the 3 forthcoming years, it is expected to grow by 7% in years 4 to 7 and after

that it is expected to grow at 6% per year in perpetuity. The required rate of return is assumed to be 9%. This information means that the DDM with three growth rates can be used to find the value of a share for this corporation.

Based on formula (2) we can express the following:

$$P_0 = \frac{D_1}{r - g_1}\left[1 - \frac{(1 + g_1)^{T_1}}{(1 + r)^{T_1}}\right] + \frac{\frac{D_{T_1+1}}{r - g_2}\left[1 - \frac{(1 + g_2)^{T_2}}{(1 + r)^{T_2}}\right]}{(1 + r)^{T_1}} + \frac{\left(\frac{D_{T_1+T_2+1}}{r - g_3}\right)}{(1 + r)^{T_1+T_2}}$$

**Table 1: The Application**

| Variable | Value | Variable | Value |
|---|---|---|---|
| $D_0$ | 2 | $r$ | 0.09 |
| $D_1$ | 2.1 | $T_1$ | 3 |
| $g_1$ | 0.05 | $T_2$ | 4 |
| $g_2$ | 0.07 | $D_{T_1+1}$ | 2.47732 |
| $g_3$ | 0.06 | $D_{T_1+T_2+1}$ | 3.21691 |
| | | $P_0$ | 71.05809 |

Notes: $D_i$ is the dividend in year $i$ and $g_j$ is the growth rate at period $j$. $T_b$ means break period at time $b$. $P_0$ represents the common stock value based on the suggested model.

As it is evident from Table 1 the present value of the underlying common stock is $71.05809. This can be a benchmark price in order to compare it with the real market price for evaluating whether the price is correctly determined or not. If the stock is not fairly priced regardless if it is overpriced or underpriced it can be used for finding a strategy that results in a gain by choosing appropriate long or short position.

## 4. Conclusions

Stock valuation is an important issue in financial markets. One model that is regularly used for this purpose is the dividend discount model (DDM). This model allows for only two different growth rates to the best knowledge. In this paper we suggest a general solution for the DDM with multiple growth rates of any potential number. The suggested solution is proved mathematically and a numerical application is also provided. The solution introduced in this paper is expected to improve on the precision of stock valuation, which might be of fundamental importance for investors, financial institutions as well as policy makers.